\documentclass[prd,aps,tightenlines,superscriptaddress,showpacs,
nofootinbib,eqsecnum,amsfonts,amsmath,twocolumn]{revtex4}
\usepackage{amssymb,amsfonts,latexsym, soul}
\usepackage{fancyhdr,color,psfrag}
\usepackage{acronym, multirow}
\usepackage[style=altlist]{glossary}
\usepackage{dcolumn}
\usepackage{subfigure, psfrag}

\newcommand{\bea}{\begin{eqnarray}} 
\newcommand{\eea}{\end{eqnarray}}
\newcommand{\ba}{\begin{array}}
\newcommand{\ea}{\end{array}}
\def\RCS$#1: #2 ${\expandafter\def\csname RCS#1\endcsname{#2}}
\RCS$Date: 2007-08-31 20:57:10 $
\RCS$Revision: 1.10 $
\usepackage{epsfig}

\newacronym{BBH}{binary black holes}{}
\newacronym{BNS}{binary neutron stars}{}
\newacronym{PBH}{primordial black holes}{}
\newacronym{SNR}{signal-to-noise ratio}{}
\newacronym{LSC}{LIGO scientific collaboration}{}
\newacronym{GRB}{gamma-ray bursts}{}
\newacronym{PSD}{power spectral sensitivity}{}
\newacronym{BHNS}{black hole - neutron star}{}
\newacronym{SPA}{stationary phase approximation}{}

\begin{document}

\title{A template bank to search for gravitational waves from 
inspiralling compact binaries: II. Phenomenological model}

\author{T. Cokelaer}
\affiliation{School of Physics and Astronomy, Cardiff University,
Cardiff CF24 3AA, UK}

\begin{abstract}
Matched filtering is used to search for gravitational waves emitted by inspiralling
compact binaries in data from ground-based interferometers. One of the key aspects of the
detection process is the deployment of a set of templates, also called a template bank,
to cover the astrophysically interesting region of the parameter space. In a companion
paper, we described the template-bank algorithm used in the analysis of LIGO data to
search for signals from non-spinning binaries made of neutron star and/or stellar-mass
 black holes; this template bank is based upon physical template families.

In this paper, we describe the phenomenological template bank that was used to search for
gravitational waves from non-spinning black hole binaries (from stellar mass formation) in
the second, third and fourth LIGO science runs. We briefly explain the design of the bank,
whose templates are based on a phenomenological detection template family. We show that
this template bank gives matches greater than 95\% with the physical template
families that are expected to be captured by the phenomenological templates.
\end{abstract}
\pacs{02.70.-c, 07.05.Kf, 95.85.Sz, 97.80.-d }

\maketitle

\section{Motivation}
The Laser Interferometer Gravitational-Wave Observatory (LIGO) detectors~\cite{LIGO} have
reached their design sensitivity curves. The fifth science run (S5) began in November 2005
and should be completed by the end of 2007, with the goal of acquiring a year's
worth of data in coincidence between the three LIGO interferometers. 
Each successive LIGO science run has witnessed improvement from both experimental and data analyst's point of
view. On the experimental side, better stationarity of the data and detector sensitivities
closer to design sensitivity curve were achieved. On the data analysis side,
the search pipeline was tuned, and new techniques were developed to reduce the background
rate while keeping detection efficiencies high. 

Among the sources of gravitational waves that ground-based detectors are sensitive
to, inspiralling compact binaries are among the most promising. Several searches
for inspiralling compact binaries in the LIGO data have been pursued: {\PBH}
\cite{LIGOS2macho,LIGOS3S4}, {\BNS}~\cite{LIGOS1iul,LIGOS2iul,LIGOS3S4}, and intermediate
mass {\BBH}~\cite{LIGOS2bbh,LIGOS3S4}. These searches used matched filtering
technique, which is the most effective and commonly used method to search
for inspiralling compact binaries.

Matched filtering computes cross correlation between the detector output and a template
waveform. If the template waveform is identical to the signal then the method is optimal,
in the sense of {\SNR}. However, in general, the template waveforms differ from the
signals. Indeed, modelizations can only approximate the exact solution of the two-body
problem. In addition, template waveforms are constructed with a subset of
the signal's parameters (e.g., the two component masses whereas eccentricity and spins
effects may be neglected). In this work, we consider only the case of non-spinning
waveforms so that the signals are entirely defined by 4 parameters: 2 mass
components, the time of arrival and the initial phase. Because the signal's
parameters are unknown, the detector output must be cross correlated with a set of
template waveforms, which is called a \textit{template bank}. While the spacing between
templates can be decreased most certainly, and this is the insurance of a {\SNR} close to
optimality, it also increases the size of the template bank (i.e., the
computational cost). The distance between templates is governed by the trade-off between
computational cost and loss in detection rate; therefore, template bank placement is a key
aspect of the detection process.

In a companion paper~\cite{squarebank}, we proposed a template bank with a minimum
match of 95\%. We assume that both template and signal were based on the same physical 
template family (precisely, the stationary phase approximation with the phase
described at 2PN order~\cite{BDI95,droz1997} ). We have shown that the template bank
could be used, effectively,
to
search for {\BNS}, {\PBH}, {\BHNS}, and {\BBH} systems. In this paper, we
consider the case of BBH systems only. 

There is a wide variety of techniques used to describe the gravitational flux and energy
generated during the late stage of the inspiralling phase (e.g., see~\cite{LB}). However,
they lead to various physical template families, and overlaps between them are not
necessarily high. In the case of heaviest systems, post-Newtonian (PN) expansion~\cite{LB}
begins to fail as the characteristic velocity $v/c$ is not close to zero anymore (e.g.,
see~\cite{BCV}). In addition, even though heavier BBH systems are accessible
each time the detector sensitivity improves in the low frequency range, BBH waveforms
remain short in the LIGO band. For instance, during the second science run
(S2)~\cite{LIGOS2bbh}, the lower cut-off frequency was set to 100~Hz, which restricted the
total mass of the search to be below $40~M_\odot$, and the longest expected signal to
last about 0.60~s. 

There exist several template families, and there is no reason to select one in
particular. A solution may be to filter the detector output with a set of template
banks, each of them associated with a different physical template families. We have shown
in
\cite{hexabank} that a unique template bank placement could be used effectively with
several template families. However, we investigate only 4 different families at 2PN
order. A different template bank might be necessary for other template families. More
importantly, the number of template families could be large, and computational cost not
manageable.  Instead of searching for {\BBH} signals using several physical template
families, a single \emph{detection template family}(DTF) was proposed  by Buonanno, Chen,
and Vallisneri~\cite{BCV} (BCV) with the goal of embedding the different physical
approximations all into a single phenomenological model. This detection template family is
known as BCV template family and has been used to  search for non-spinning BBH signals in
LIGO data~\cite{LIGOS2bbh,LIGOS3S4}.


In this paper, we do not intend to compare a search that uses BCV
templates and a search based upon physical template family. Our main goal is to describe
the BCV template bank that was developed and used to search for stellar-mass BBH signals
in the second (S2), third (S3), and fourth (S4) LIGO science runs~\cite{LIGOS2bbh,
LIGOS3S4}. In section~\ref{sec:introduction}, we briefly
discuss the template parameters and the filtering process related to BCV
templates. In section~\ref{sec:bankdesign}, we describe the BCV template bank, and the
the spacing between templates. In
section~\ref{sec:simulations}, we test and validate the proposed template bank
with exhaustive simulated injections. Finally, in section~\ref{sec:conclusion},
we summarize the results.

\section{The BCV template family}\label{sec:introduction}
The detection template family that was proposed in~\cite{BCV} is built directly
from the Fourier transform~\cite{BSD95} of gravitational-wave signals by writing the
amplitude
and phase as polynomials in the gravitational-wave frequency law that appear in
the stationary phase approximation~\cite{SD91}. In the frequency domain, the BCV
templates
are defined to be
\begin{equation}\label{eq:template}
h\left(f\right) = \mathcal{A}(f) e^{i
\psi\left(f\right)}\,,
\end{equation}
where
\begin{eqnarray}
\label{eq:amplitude}\mathcal{A}(f) & =& f^{-7/6}\left( 1 - \alpha f^{2/3}\right)
\theta\left( f_{\rm cut} - f \right),\,\\
\label{eq:phase}\psi(f) &=& 2\pi f t_0 + \phi_0 + f^{-5/3} \sum_{k=0}^{N}
\psi_{k}f^{k/3}\,.
\end{eqnarray}
The parameters $t_0$ and $\phi_0$ are the standard time of arrival and initial phase of
the gravitational wave signal. The parameter $\alpha$ is a shape parameter
introduced to capture post-Newtonian amplitude corrections. Because various
models predict different terminating frequencies, an ending cut-off frequency
$f_{\rm cut}$ is introduced. In the amplitude expression, the waveform is
multiplied by a Heaviside step function, $\theta\left( f_{\rm cut} - f \right)$.
In the right
hand side of equation~\ref{eq:phase}, we use only two parameters $\psi_0$
and $\psi_3$, which  suffices to obtain a high match with most of the PN models
\cite{BCV}. The symbol $\psi_3$  here is the same as the symbol
$\psi_{3/2}$ in~\cite{BCV}. 

The $\psi_k$ parameters are the phase parameters of the phenomenological waveform, which
cannot be directly linked to the physical mass parameters; the BCV
templates are made for detection, not for parameter estimation. Nevertheless, a
good approximation (for low masses) of the chirp mass $\mathcal{M}$  is given
by  
\begin{equation}\label{eq:chirpmass}
\psi_0 \approx \frac{3}{128} \left( \frac{1}{\pi \mathcal{M}}\right)^{5/3}\,.
\end{equation}
In section~\ref{sec:discussion}, we investigate the range of validity of this
relation.

The filtering of a data set using BCV templates is not as trivial as the one that
uses physical template families. Indeed, the BCV filtering implies a search in six
dimensions ($\psi_0$, $\psi_3$, $\alpha$, $\phi_0$, $t_0$, and $f_{\rm cut}$).
The {\SNR} can be analytically maximized over $\alpha$, $\phi_0$ and
$t_0$, which reduces the search to three dimensions only. 
the
maximization over  $\alpha$ and $\phi_0$ is not. In order to perform the filtering
and the maximization over $\alpha$ and $\phi_0$, we
need to construct orthonormal basis vectors
$\left\{\hat{h}_k\right\}_{k=1,..,4}$ for the 4-dimensional linear subspace of templates
with $\phi_0 \in \left[0,2\pi\right)$ and $\alpha \in \left(-\infty,\infty\right)$, and
we want the
basis vectors to satisfy $\left\langle \hat{h}_i \Big| \hat{h}_j\right\rangle=\delta_{ij}$
(see the
Appendix for details). The SNR before
maximization is given by 
\begin{eqnarray}\label{eq:rho}
\rho &=& x_1 \cos{\omega} \cos{\phi_0} +x_2 \sin{\omega} \cos{\phi_0} \\\nonumber
&+&x_3 \cos{\omega} \sin{\phi_0}  + x_1 \sin{\omega} \sin{\phi_0}\,,
\end{eqnarray}
where $x_i=\left\langle s \Big| \hat{h}_i\right\rangle$, and $s$ is the data
to be filtered
\footnote{The expression of the SNR shows that the expected rate of false alarm follows
a chi-square distribution with 4 degrees of freedom instead of 2 in the case of physical
template families}. The parameter $\omega$ is a function of $\alpha$ (see
equation~\ref{eq:omega} and \ref{eq:alphaomega}).

The {\SNR} $\rho$ can be maximized over $\phi_0$ and $\omega$($\alpha)$. In
\cite{LIGOS2bbh}, the maximization is done over the two new parameters
$A=\omega+\phi_0$ and $B=\omega-\phi_0$. The maximized
{\SNR} (independent of $\alpha$ and $\phi_0$), is denoted $\rho_U$, and is given by  
\begin{equation}\label{eq:rhou}
\rho_U = \frac{1}{\sqrt{2}}\left( \sqrt{V_0 +\sqrt{V_1^2+V_2^2} }\right)\,,
\end{equation}
where $V_k$ are function of $x_i$ (see Appendix and equations
\ref{eq:V1}, \ref{eq:V2} and \ref{eq:V3}).

The SNR provided in equation~\ref{eq:rhou} is the \textit{unconstrained} SNR that is
independent of any constraint on the range of the  parameter $\alpha_F^U=\alpha f^{2/3}$
(again, here the index $U$ represents the unconstrained case, and we shall use $C$ for
the constrained case). Yet, in~\cite{BCV}, the authors suggested that the parameter
$\alpha_F^U$ should be restricted to the range $[0, 1]$. Indeed when $\alpha_F^U > 1$, the
amplitude in equation~\ref{eq:amplitude} becomes negative, which corresponds to unphysical
waveforms. Moreover, when $\alpha_F^U < 0$, the amplitude factor can substantially deviate
from the predictions of PN theory.

In S2~\cite{LIGOS2bbh}, many accidental triggers were found with
$\alpha_F^U > 1$, and the calculation of the {\SNR} was unconstrained (as in
equation~\ref{eq:rhou}) leading to a high false alarm rate, which was decreased, a
posteriori, by removing all triggers for which $\alpha_F^U > 1$ (without decreasing the
detection efficiency). Nevertheless, triggers that verified $\alpha_F^U < 0$ were kept
because the false alarm rate did not decrease significantly when this selection was
applied.

In S3 and S4, the search for BBH systems deployed a filtering that
takes $\alpha_F^U$ value into account, by using a maximization of
equation~\ref{eq:rho} that leads to a \textit{constrained} SNR denoted $\rho_C$. 
The expression for the constrained SNR depends now on the value of $\alpha_F^U$. We have
$\alpha_F^C=\alpha_F^U$ if $0
\leq \alpha_F^U \leq 1$, and no constraint is applied (i.e., $\rho_C = \rho_U$).
However,
if $\alpha_F^U < 0$ or $\alpha_F^U > 1$, then  a constrained SNR is used so that the final
$\alpha_F^C$ parameter is 0 and 1, respectively, and $\rho_C \leq \rho_U$. The expressions
of the constrained SNR are provided in the Appendix. 

Let us make the point clear. In the contrained-SNR case, as
explained above, $\alpha_F^C=\alpha_F^U$ if $0 < \alpha_F^U < 1$ only. Otherwise
$\alpha_F$ takes
only two values (0 or 1) but we know $\alpha_F^U$ (since it is the condition to
apply the constraint or not). 

It is worth noticing that for the study that follows, we always use a constrained SNR 
but using an unconstrained SNR should not significantly change the results of our
simulations and/or template bank placement. Indeed, simulated injections are generated
with physical template families for which we do not expect $\alpha_F$ to be unphysical
(i.e., outside $[0-1]$), as  we shall see in section~\ref{sec:conclusion}. Using a
constrained SNR has an important impact when dealing with real analysis, where
most
of the accidental triggers have $\alpha_F$ between $\	\left(-\infty\,, \infty\right)$
(and
therefore in $(-\infty\,, 0[\; \cup \;]1\,, \infty)$ as well, where $\rho_C <
\rho_U$). Consequently, in general, for a given threshold $\lambda$, the SNR of
accidental triggers  have $\rho_C <  \rho_U$, and the rate is therefore 
lower with respect to a search with unconstrained SNR. The number of triggers that
needs to be stored is lower by an order of magnitude. Nevertheless, the final rate of
triggers between the two methods may be equivalent because of
\textit{a posteriori} cuts on $\alpha_F$ when an unconstrained SNR is used as in
\cite{LIGOS2bbh}.

\section{BCV template bank design}\label{sec:bankdesign}
Template bank placement has been investigated in several papers 
\cite{Owen96,OwenSathyaprakash98,buskulic,squarebank,hexabank,flatbank} in the context of
physical template families. We refer the reader to the established literature in
this subject area.

\subsection{Metric Computation in $\psi_0\textrm{--}\psi_3$
Plane}\label{sec:psi0psi3}

In the case of BCV templates, the mismatch metric $g_{ij}$~\cite{Owen96} is
known (see Appendix), and is constant over the entire $\psi_0\textrm{--}\psi_3$ parameter
space. Nonetheless, the metric components are strongly
related to the lower cut-off frequency of the search, which affects the moments
used to calculate the metric (see equation~\ref{eq:moments}). The moment computation
also depends on the $\alpha$ parameter, as discussed later. For now, let us suppose that
the
moments are fixed.

Because the metric is constant, the placement of templates on the
$\psi_0\textrm{--}\psi_3$ parameter space is straightforward. In the first search for BBH
signals~\cite{LIGOS2bbh}, the template placement used a square
lattice, and templates were placed parallel to the $\psi_0$ axes.  In S3 and S4 BBH
searches, an optimal placement was used (hexagonal lattice), which reduced the
requested computing resources (and trigger rate) by 30\% with respect to S2. In
this paper, we only consider tests related to the hexagonal lattice case. In S3 and S4
BBH searches, we placed the templates parallel to the first eigenvector rather than
parallel to the $\psi_0$ axis. 

The target waveforms are BBH systems for which the lowest component mass is set to 
$3~M_\odot$ and the highest component mass is defined by the detector lower
cut-off frequency (up to $80~M_\odot$ in S4). Simulations show that to
detect such target waveforms, the range of phenomenological parameters
should be set to $\psi_0 \in [10000,\;550000]~{\rm
Hz}^{5/3}$ and $\psi_3 \in [-5000,\;-10]~{\rm Hz}^{2/3}$. As explained in
section~\ref{sec:polygon}, if we search for BBH systems only, a significant fraction of
those templates are not needed and can be removed from the template bank.

\subsection{$\alpha_B$-dependence}
\begin{figure}[ht]
\centering
{\includegraphics[width=0.42\textwidth]{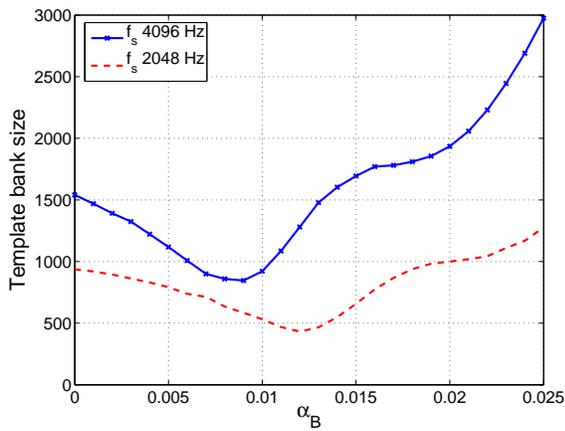}}
\caption{The size of the template bank function of the $\alpha_B$ parameter. The two
curves show the template bank size versus $\alpha_B$ parameter for two values of
sampling frequency. The two curves show the same pattern with a drop around
$\alpha_B=10^{-2}$, where template bank sizes are twice as low as compared to 
$\alpha_B=0$ or $\alpha_B=2.5\times\,10^{-2}$. This evolution of the template bank size is
directly linked to the moment computation (see equation~\ref{eq:moments}), where the
parameter $\alpha_B$ is used. Efficiency of a template bank is not strongly
related to this parameter (see figure~\ref{fig:alphaB_eff}) so we choose a value
that corresponds to the smallest bank size. In real analysis, we use 2048~Hz and
for simplicity a value of $\alpha_B=10^{-2}$ was chosen. In this example, we
used the same simulation parameters as in section~\ref{sec:example1}.}
\label{fig:alphaB}
\end{figure}

\begin{figure}[ht]
\centering
{\includegraphics[width=0.45\textwidth]{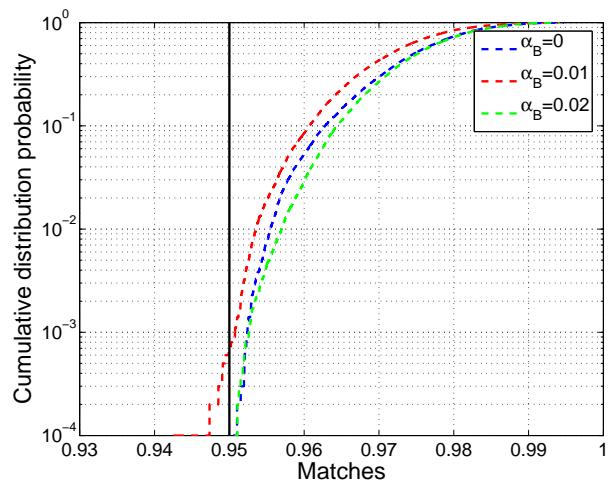}}
\caption{Template bank efficiencies versus $\alpha_B$ parameter. The $\alpha_B$
parameter does not  significantly affect the matches. Most of them are above
 the minimal match of 95\%, and more importantly the three distributions are
close to each other. In this example, we used the same simulation parameters as
in section~\ref{sec:example1}, and EOB injections.}
\label{fig:alphaB_eff}
\end{figure}

The moments used to estimate the metric components strongly depend on the
parameter $\alpha$. We refer to this parameter as $\alpha_B$ to
differentiate from the $\alpha$ parameter (or equivalently from $\alpha_F$) that is used
in the
filtering process.  As shown in figure~\ref{fig:alphaB},
the number of templates changes significantly when $\alpha_B$ varies. 
There is a drop in the number of templates around $\alpha_B=10^{-2}$.
We want to minimize the template bank size but  we also need to
consider the \textit{efficiency} of the bank as defined in
\cite{squarebank, hexabank}, and choose $\alpha_B$ appropriately.
Indeed, we expect the efficiency of the template bank to be also affected by
this
parameter.  We performed simulated injections so as to test the efficiency of the
template bank for various values of $\alpha_B$. Results are summarized in
figure~\ref{fig:alphaB_eff} for three typical values of $\alpha_B$. Because
efficiencies are very similar, we decided to use an $\alpha_B$ parameter such
that the number  of templates is close to a minimum, that is $10^{-2}$. In all
the following simulations and LIGO searches, $\alpha_B=10^{-2}$.

\subsection{Template Bank using ending frequency layers}

Starting from each template that is placed in the $\psi_0\textrm{--}\psi_3$
plane, we
need to lay templates along the third dimension, which is the ending cut-off
frequency $f_{\rm cut}$ of the template. Because the mismatch is first order
in $\Delta f_{\rm cut}$~\cite{BCV}, it cannot be described by  a metric. 

Using an exact formula,~\cite{BCV} proposed to lay templates with different $f_{\rm cut}$
values between $f_{\rm min}$ and $f_{\rm end}=f_{\rm Nyquist}$ that depend on
the region searched for. We populate the $f_{\rm cut}$ dimension as follows. First,
we estimate the frequency of the last stable orbit which we refer to as $f_{\rm
min}$, and the frequency at the light ring which we refer to as $f_{\rm max}$.
Between $f_{\rm min}$ and $f_{\rm max}$, we place $N_{\rm cut}$ layers of templates
with the ending frequency chosen at equal distance between  $f_{\rm
min}$ and $f_{\rm max}$. The frequency at the last stable orbit and light ring are
defined in terms of the total mass ($f_{\rm min} = 1/( M \pi 6^{3/2})$, $f_{\rm
max}=1(M \pi 3^{3/2})$). The total mass
is computed for each template using an empirical expression similar to
equation~\ref{eq:chirpmass}: $M \approx -\psi_3/(32 \pi^2\psi_0)$. This
expression is an approximation. It underestimates the total mass for low mass range,
however, it is suitable for the wide range of mass that we are interested in: the final
template bank gives high match with the various physical template families, as shown in
section~\ref{sec:simulations}. In all our simulations and searches, we set the minimal
match $(MM)$~\cite {Owen96} to $95\%$, so there is no guarantee that the relation between
$M$ and $\psi_{0,3}$ is suitable for minimal matches far from 95\%.

\begin{figure}[ht]
\centering
{\includegraphics[width=0.45\textwidth]{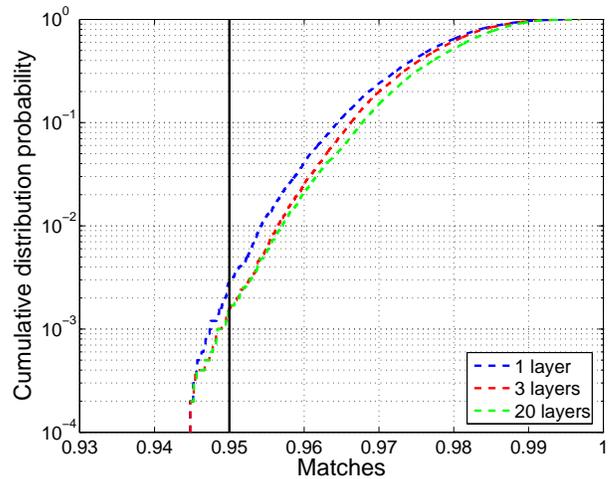}}
\caption{Template bank efficiencies versus number of layers, $N_{\rm cut}$ ,  in
the $f_{\rm cut}$ dimension. With the current template bank design, $N_{\rm cut}$
does not affect the matches significantly. The cumulative distribution of matches over
10,000 simulations shows only small differences between 3 and 20 layers. Even the results
obtained with 1 layer are not that far from $N_{\rm cut}=3$. In
our real analysis and simulations, we used $N_{\rm cut}=3$. In this example, we used the
same simulation parameters as in section~\ref{sec:example1}, and EOB simulated
injections.}
\label{fig:flayers}
\end{figure}

\subsection{Polygon Fit}\label{sec:polygon}

The boundaries of the template bank are defined by the ranges of the parameters $\psi_i$ 
and the span of the cut-off frequency $f_{\rm cut}$ in such a way that BBH systems
with component mass as low as
$3~M_\odot$ are detectable. The $\psi_i$ ranges provided in section~\ref{sec:psi0psi3}
cover
a squared area
that is actually too wide: a significant fraction of the templates are not
targeting the BBH
systems we are searching for. Therefore, in order to reduce the template bank
size and optimize our searches, we introduce an extra procedure that selects the
pertinent templates only. This procedure is known as a \textit{polygon fit} and
works as follows. First, we create a BCV template bank with the range of
$\psi_0$ and $\psi_3$ parameters as large as possible, and for our purpose, as
quoted in section~\ref{sec:psi0psi3}. This choice of ranges allows us to not only detect
BBH systems but also {\BHNS} systems. Since, we
focus on
the BBH systems only, we perform many BBH simulated injections and filter them with the
template bank that has been created. For each injection, we keep the $\psi_0$ and $\psi_3$
parameters of the template that gives the best match. We gather all the final pairs of
$\psi_0$, $\psi_3$ parameters, and
superpose them on top of the original template bank. It appears that only about
a third of the templates are required to detect BBH systems with a high match.
This sub-set of templates can be used to define a polygon area
that enclose all of them. The resulting polygon area defines the boundaries of our new
template bank and results in a template bank three times smaller than the original one. 

\begin{figure}[ht]
\centering
{\includegraphics[width=0.45\textwidth]{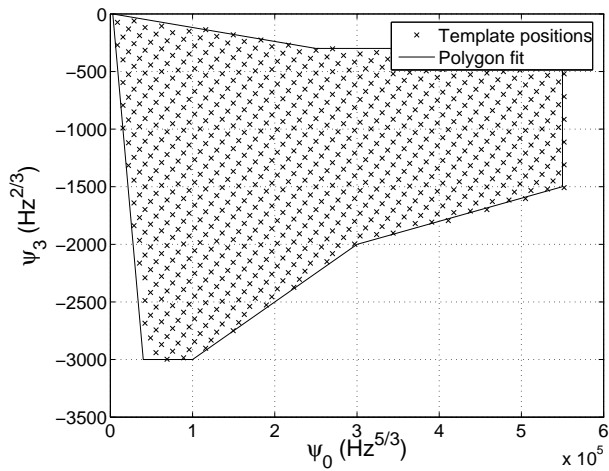}}
\caption{Example of parameter space and template bank placement. This plot
shows a projection of the templates onto the $\psi_0/\psi_3$ plane. Simulations
and equation~\ref{eq:chirpmass} gives an estimation of the mapping between the
phenomenological parameters and
the chirp mass of the simulated injections. Low mass systems such
as a
$(3,3)M_\odot$ are in the RHS, high mass systems lie on the LHS and asymmetric systems in
the bottom left corner. In this example, we used the same parameters as in
section~\ref{sec:example2}.}
\label{fig:polygon}
\end{figure}

In
figure \ref{fig:polygon}, we show such a template bank that is within
the boundary of a polygon constructed with our simulated injections. The
coordinates of this polygon are chosen empirically. For safety, the boundaries are chosen
loosely, therefore the template bank
has also the ability to detect non-spinning {\BHNS}. It is worth noticing that
with this template bank designed to detect {\BBH} many {\BHNS} systems are found with
a match greater than the requested $MM$ (See section~\ref{sec:example3}).

\section{Simulations}\label{sec:simulations}
In  the following simulations, we fix the sampling frequency to
$2048$~Hz, $\alpha_B=10^{-2}$, $N_{\rm cut}=3$, the $\psi_0$ and $\psi_3$ ranges are
provided in section~\ref{sec:psi0psi3} and a polygon fit as in figure~\ref{fig:polygon} is
used. The
simulated injections are based on several physical template
families that are labelled EOB, PadeT1, TaylorT1, and TaylorT3
\cite{EOB1,EOB2,EOB3,Standard,DIS} with the phase expressed 
at 2PN order (see~\cite{hexabank} for more details). The population of simulated
injections has a uniform total mass. Although this choice is not based on any
astronomical observation, it is convenient to estimate the efficiency of our template
banks. We use a
noise model that mimics the design  sensitivity curve of
initial LIGO (see~\cite{hexabank,BSD95}), and the minimal match is $MM=95\%$.
We performed 2 simulations that are closely related to the third and fourth LIGO
science run's BBH searches~\cite{LIGOS3S4}. 

\subsection{Example 1}\label{sec:example1}
\begin{figure}[th]
\centering
{\includegraphics[width=0.45\textwidth]{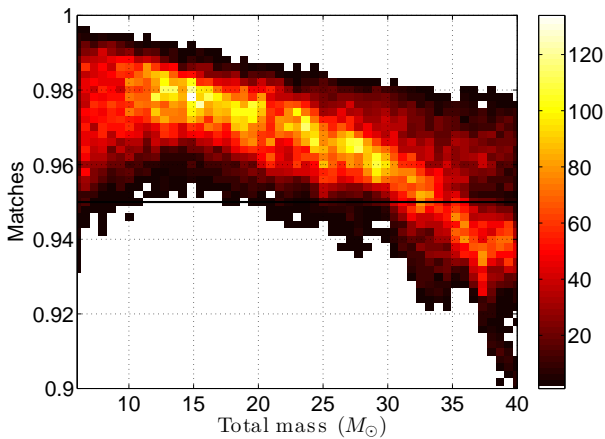}}
\caption{Distribution of the efficiency versus the total mass. The simulation consisted
of $N_s=40,000$ injections. The lower cut-off frequency for the injections and the BCV
templates was set to 70~Hz, as in S3 BBH search.}
\label{fig:s3}
\end{figure}

The first set of simulations uses a lower cut-off frequency of $70$~Hz, as in
S3 BBH search~\cite{LIGOS3S4}. The maximal total mass of
the simulated injections is set to $40~M_\odot$ and therefore the largest component mass
to $37~M_\odot$. The template bank has 531 templates. The results are summarized in
figure~\ref{fig:s3} which shows the efficiency of the template bank versus the total mass.
There are a few injections found with a match as low
as 93\% for total mass $M < 6.5~M_\odot$. Closer inspection shows that several
issues are linked to this feature. First, we used a sampling
frequency of 2048~Hz, which reduces the template bank size by
$\approx50\%$ as compared to a sampling of 4096~Hz. Second, we set
$\alpha_B=10^{-2}$, which reduces the template bank size by $\approx50\%$ as
compared to $\alpha_B=0$. Finally, the number of layers, $N_{\rm cut}$,
is
 limited to 3. Therefore, this tuning significantly reduced the
template bank size with the cost
of losing about only 1 to 2\% SNR for a small fraction of the parameter space
considered. From $M=6.5~M_\odot$ to about $M=20~M_\odot$, matches are above
95\%. In the high mass range, a large fraction of the simulated
injections are found below the minimal match (but larger than 90\%): 20\% in the case
of TaylorT1, TaylorT3, and PadeT1 models, and only 0.1\% in the case of EOB
injections. This effect is expected because the lower cut-off frequency is high, and
therefore many of the high mass systems considered are very short (i.e., less than
100~ms). Because the final frequency of the EOB signals goes up to the light ring, the
matches are larger than in the case of TaylorT1, TaylorT3, and PadeT1 approximants, whose
last frequencies stop at the last stable orbit.

\subsection{Example 2}\label{sec:example2}
\begin{figure}[tbh]
\centering
{\includegraphics[width=0.45\textwidth]{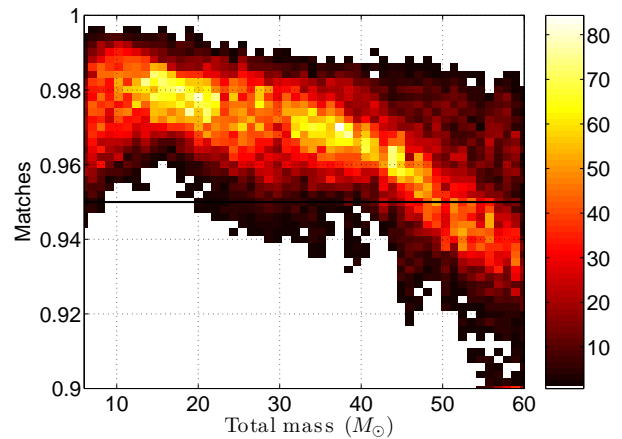}}
\caption{Distribution of the efficiency versus the total mass. The simulation
consisted in $N_s=40,000$  injections. The lower cut-off frequency of the injections and
the BCV templates is 50~Hz, as in S4 BBH search. For convenience (same
scale as in figure~\ref{fig:s3}), we do not show simulated injections with total
mass above $60~M_\odot$ and matches below 90\% (see text for details).}
\label{fig:s4}
\end{figure}

The second set of simulations uses a lower cut-off frequency of $50$~Hz, as in
S4 BBH search~\cite{LIGOS3S4}. The maximal total mass of the simulated injections
is set to $80~M_\odot$ and therefore the largest component mass to $77~M_\odot$. The
template bank has 1609 templates. The results
are summarized in figure~\ref{fig:s4}. Up to $M\approx40~M_\odot$, most of the
injected simulations are recovered with matches above 95\%. However, a small fraction
is found with matches below 95\%,
which represent 0.1\% of the EOB, PadeT1, and TaylorT1  injections, and
3\% of the TaylorT3  injections. In the high mass
region (up to $60~M_\odot$),  20\% of the  injections are
below the required minimal match for the TaylorT1, TaylorT3, and PadeT1
 injections, and only 0.5\% of the EOB  injections. If we
consider  injections with total mass from 60 to $80~M_\odot$, almost 10\% of EOB
are below the minimal match (but above 92\%). As for other models, matches drop
quickly towards zero down to 40\%, which is due to shorter and shorter duration
of the  injected waveforms.

\subsection{Example 3}\label{sec:example3}
As stated in section~\ref{sec:polygon}, although the template bank is designed to target
BBH
systems, it has the ability to detect some BHNS systems as well. The goal of this
third simulation is to demonstrate that indeed many BHNS systems are detectable with a
high
match by using the template designed to search for BBH systems in S3 and S4 data
sets. 

\begin{figure}[tbh]
\centering
{\includegraphics[width=0.45\textwidth]{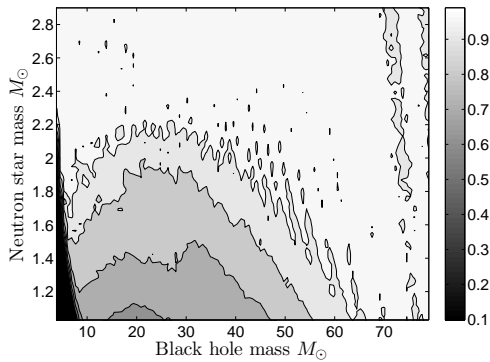}}
\caption{Matches between the BCV template bank used to search for BBH systems and
BHNS injections.  The simulation
consisted in $N_s=100,000$  injections. We found that 77\% of the BHNS injections have
 a match larger than 90\%. See
the text for more details. }
\label{fig:bhns}
\end{figure}

The parameters used are exactly the same as in the second example. The maximal
total mass of the simulated injections is set to $80~M_\odot$, the largest component to
$79~M_\odot$, and the lowest component mass is set to $1~M_\odot$. We impose the
systems to be {\BHNS} only (the
mass of the neutron star is less than $3~M_\odot$, and the mass of the black hole is
larger than $3~M_\odot$). The
template bank is identical to the second simulation (1609 templates). The results are
summarized in figure~\ref{fig:bhns}, where we plot matches as a function of the two
component masses. We found that 60\% of the BHNS injections are recovered with the
match
larger than 95\%, 77\% with the match larger than 90\%, and 98\% with the match greater
than
50\%. Therefore, using the same bank as in S3 and S4 searches, whose boundaries
resulting from the polygon fit were deliberately chosen to be slightly wider than
necessary, we can detect a significant fraction of the BHNS systems. 
It is also clear from the figure that the lightest systems have a very low match. This
was expected since the template bank aimed at detecting systems whose total mass
is greater than $6 M_\odot$, as defined by the maximum of the $\psi_0$ range. 

We performed a second test where the polygon fit is not applied anymore. The
template bank is then much larger with 4635 templates but we found that 78\% of the BHNS
injections are recovered with a match larger than 95\%, 94\%  with a match larger than
90\%, and 98\% with a match greater than 50\%. The size of such a template bank is
comparable to a template bank that uses physical template families (e.g., with the same
parameters as above, a hexagonal placement for physical template families~\cite{hexabank}
that covers a parameter space from 1 to 80 solar mass has about 3000 templates if we
exclude the templates for which both component mass are below 3 $M_\odot$). 

The events which are found with a low match (say, 60\% or lower) correspond to low
mass systems where the neutron star's mass is less than $2.5~M_\odot$ and the BH's
mass less than $7~M_\odot$ which can be taken care of by increasing the range of
$\psi_0$.

\begin{figure}[tbh]
\centering
{\includegraphics[width=0.45\textwidth]{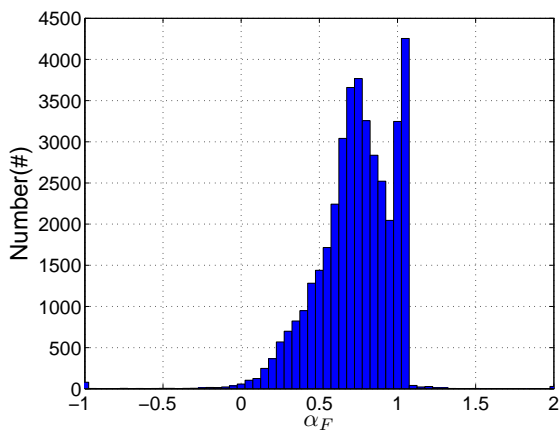}}
\caption{Distribution of the $\alpha_F^U$ parameter corresponding to simulations
made in section~\ref{sec:example2}. Most of the found simulated injections have an
$\alpha_F^U$ value between zero and unity. However, a significant fraction are
distributed around $\alpha_F^U=1$. Those triggers correspond to $M>60~M_\odot$, for which
waveforms cannot be differentiated from a transient noise (short duration).}
\label{fig:alphaF}
\end{figure}

\begin{figure}[tbh]
\centering
{\includegraphics[width=0.45\textwidth]{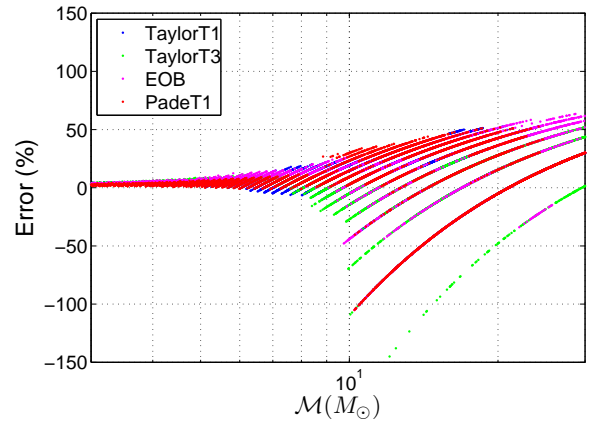}}
\caption{Errors in chirp mass estimation corresponding to simulations made in
section~\ref{sec:example2}. Errors increase significantly for $\mathcal{M}$ greater than 8
solar mass with errors larger than 50\%.} 
\label{fig:error}
\end{figure}

\subsection{Discussions}\label{sec:discussion}
In this section, we use the results of section \ref{sec:example2} to check 
  (i) the range of validity of equation~\ref{eq:chirpmass}, which gives an
estimation of the chirp mass, and (ii)  the regime of constrained SNR (i.e.,
the value of $\alpha_F$).

Although we use a constrained SNR, we kept track of the value of $\alpha_F^U$ before
the maximization. We plot the distribution of $\alpha_F^U$ in figure
\ref{fig:alphaF}. About 83\% of the injections were found with a $\alpha_F^U$
value in the range ]0,1[. Therefore, as stated in section~\ref{sec:introduction}, the
results obtained with the constrained SNR are very similar to what we would have obtained
if we had used the unconstrained SNR. The distribution has a first peak around 0.7 and  a
second peak in the range [1, 1.1], which correspond to about 15\% of the injections; it
corresponds to total mass above $60~M_\odot$. 

In figure~\ref{fig:error}, we plot the errors on the chirp mass (i.e.,
$(\mathcal{M}_i-\mathcal{M}_e)/\mathcal{M}_i$, where $i$ stands for injected, and $e$ for
 estimated).  We used equation~\ref{eq:chirpmass} to estimate the chirp mass. The 
errors are within 10\% for BBH systems when the chirp mass is below
$\approx 8~M_\odot$. However, errors increase significantly when $\mathcal{M} \gtrsim
8~M_\odot$ because (i) parameter estimation of high mass BBH systems is intrinsically
weak, even for physical template families and (ii) BCV templates are known to
be detection template families that are not suitable for parameter estimation.

\section{Conclusions}\label{sec:conclusion}

The BCV template bank that we described in this paper was used to search for BBH
systems in the S2, S3 and S4 LIGO data sets. We described the significant
improvements that were made between the S2 search and the
S3/S4 searches: $\alpha_B$ tuning, hexagonal lattice, and
polygon fit. These improvements reduce
the template bank size by an order of magnitude, while keeping the efficiency
higher than 95\% for most of the BBH systems considered. Consequently, despite reducing
the lower cut-off frequency from 100Hz to 50Hz between the second and
the fourth science runs, the template bank size remained similar.

A principal motivation for the construction of a detection template bank was
to use a single template
bank instead of several physical template families. The template bank size is
therefore an important aspect of a BCV search, and we have shown in this work how
the number of templates can be optimized to search for BBH. Remarkably, the same template
bank has a high match with a wide range of BHNS. 


More importantly, the BCV template bank was designed to search for BBH systems
in the context of the S2 LIGO search. That is, for a lower cut-off frequency of 
100~Hz for which most of the target waveforms are short duration waveforms.
However, LIGO detectors improved and are still improving at low frequency,
making the waveforms longer. The advantage of using a BCV template to search
for systems as low as $3~M_\odot$ is no longer evident, especially
considering the absence of a well defined $\chi^2$ test for phenomenological
templates. Therefore if it were to be used, the author
thinks that a BCV template bank should be used to search for a mass range starting at a
higher value, such as $10$ or $20~M_\odot$.

\begin{acknowledgments}
We would like to acknowledge many useful discussions with members of the LIGO
Scientific Collaboration, in particular of the LSC-Virgo Compact Binary Coalescence
working group, which were critical in the formulation of the results described in this
paper. This work has been supported in part by Particle Physics and Astronomy Research
Council, UK, grant PP/B500731. This paper has LIGO Document Number LIGO-P070089-01-Z.
\end{acknowledgments}

\begin{appendix}

\section{Filtering, $\alpha$-maximization, and constrained SNR}
We define the inner product as follows
\begin{equation}
 \left< h_1\Big|h_2\right> = 4 Re \left( \int_0^\infty \frac{h_1(f)
h_2^*(f)}{S_h(f)}df\right)\,,
\end{equation}
where $S_h(f)$ is the one-sided noise power spectral density. 

\subsection{Filtering}
The BCV templates in the frequency domain are  defined by equation~\ref{eq:template}. The
amplitude part of a BCV template $\mathcal{A}(f)$  can be decomposed into
linear combinations of $f^{-7/6}$ and $f^{-1/2}$. These expressions can be used
to construct an orthonormal basis $\{\hat{h}_k\}_{k=1..4}$. We want the basis vectors to
satisfy 
\begin{equation}
\left< \hat{h}_i \Big| \hat{h}_j\right> = \delta_{ij}.
\end{equation}
First, we construct two real functions  $\mathcal{A}_1(f)$ and
$\mathcal{A}_2(f)$ that satisfy $\left< A_i\big| A_j\right> = \delta_{ij}$. Then,
we define $\hat{h}_{1,2}(f)=\mathcal{A}_{1,2}(f)e^{i(\psi-\phi_0)},\hat{h}_{3,4}
(f)= i \mathcal{A}_{1,2}(f)e^{i(\psi-\phi_0)}$ which will give  $\left< \hat{h}_i\big|
\hat{h}_j\right> = \delta_{ij}$, and the desired basis, $\{\hat{h}_k\}$. $\psi$ is the
phase of the signal, as defined in equation~\ref{eq:phase}, and $\phi_0$ the initial
phase that we want to maximize. We can
choose the following basis functions

\begin{equation}
\left[	
	\begin{array}{cc} 
		\mathcal{A}_1(f) \\
		\mathcal{A}_2(f) 
	\end{array}
\right]
=
\left[
	\begin{array}{cc} 
		a_{11} & 0 \\
		a_{21} & a_{22} 
	\end{array}
\right]
\left[
	\begin{array}{cc} 
		f^{-7/6} \\
		f^{-1/2}
	\end{array}
\right],
\end{equation}
where the normalization factor are given by 
\begin{eqnarray}\label{eq:normalisation}
a_{11} = I^{-1/2}_{7/3},
a_{21} = - \frac{I_{5/3}}{I_{7/3}} \left( I_1 -
\frac{I^2_{5/3}}{I_{7/3}}\right)^{-1/2}, \\{\rm and \;}
a_{22} = \left(I_1 - \frac{I^2_{5/3}}{I_{7/3}} \right)^{-1/2}\,,
\end{eqnarray}
and the integrals $I_k$ are defined by
\begin{equation}
 I_k = 4 \int_0^{f_{\rm cut}} \frac{f^{-k}}{S_h(f)}df.
\end{equation}

The normalized template can be parametrized using the orbital phase
$\phi_0$ and an angle $\omega$
\begin{eqnarray}
\hat{h}(\theta, \omega; f) &=& \\\nonumber
&&\hat{h}_1(f) \cos \omega \cos \phi_0 +
\hat{h}_2(f) \sin \omega \cos \phi_0 \\\nonumber
&+& \hat{h}_3(f) \cos \omega \sin \phi_0 +
\hat{h}_4(f) \sin \omega \sin \phi_0
\end{eqnarray}
where $w$ is related to $\alpha$ by (see~\cite{LIGOS2bbh})
\begin{equation}\label{eq:omega}
\tan{\omega} = - \frac{a_{11} \alpha}{a_{22}+a_{21} \alpha}\,, 
\end{equation}
which can be inverted to get $\alpha$
\begin{equation}\label{eq:alphaomega}
\alpha =  -\frac{a_{22} \tan{\omega}}{ a_{21} \tan{\omega} + a_{11}}. 
\end{equation}
It follows that for any given signal $s$, the overlap is 
\begin{eqnarray}\label{eq:rho_app}
\rho &=& \left<s\Big|\hat{h}\right> \\\nonumber
&=&x_1 \cos
\omega \cos \phi_0 +
x_2 \sin \omega \cos \phi_0 \\\nonumber
&+& x_3 \cos \omega \sin \phi_0 +
x_4 \sin \omega \sin \phi_0,\,
\end{eqnarray}
where $x_i = \left<s \Big| \hat{h}_i\right>$. We can then maximized over $\omega$
(i.e., $\alpha)$, and $\phi_0$ without any constraint on the $\alpha$ parameter, which
leads to the unconstrained SNR given by 
\begin{equation}\label{eq:rhou_app}
\rho_U =\frac{1}{\sqrt{2}}\sqrt{\left(V_0 +\sqrt{V_1^2+V_2^2}\right)}
\end{equation}
where
\begin{eqnarray}
V_0 &=& x_1^2 +x_2^2 +x_3^2+x_4^2\,,\\\label{eq:V1}
V_1 &=& x_1^2 +x_3^2 -x_2^2-x_4^2\,,\\\label{eq:V2}
V_2 &=& 2(x_1x_2+x_3x_4 ).\label{eq:V3}
\end{eqnarray}

The values of $\omega$ and $\phi_0$ that maximize $\rho_U$ are provided in
\cite{LIGOS2bbh} as function of the $x_i$. We reformulate $\omega^{\rm max}$ using the
$V_i$ and found this simple expression:
\begin{equation}\label{eq:wmax}
 \tan{2\omega^{\rm max}} = \frac{V_2}{V_1}\,. 
\end{equation}
It is then straightforward to obtain $\alpha^{\rm max}$ using equation
\ref{eq:alphaomega}, and $\alpha_F^U = \alpha^{\rm max} f^{2/3}$.

\subsection{constrained and unconstrained SNRs}
Starting from equation~\ref{eq:rho_app}, we can derive a constrained SNR $\rho_C$
that depends upon the value of the parameter $\alpha_F$. Therefore, we need to
maximize equation~\ref{eq:rho_app} over the parameter $\phi_0$ only. This maximization
gives

\begin{equation}\label{eq:rho_phi0}
\rho(\omega) =\max_{\omega}  \frac{1}{\sqrt{2}} \sqrt{V_0 + V_1 \cos 2\omega
+V_2 \sin 2\omega}\,.
\end{equation}

If $\alpha_F <0$, we want to use a SNR calculation for which $\omega=0$, which
means $\alpha_F^C=\alpha=0$. Therefore, the constrained SNR is 
\begin{equation}\label{eq:rhoc0}
\rho_C^0 = \frac{1}{\sqrt{2}} \sqrt{V_0 + V_1}\,.
\end{equation}

If $\alpha_F >1$, we want to use a SNR calculation for which $\omega=\omega_{\rm
max}$, which means that $\alpha_F^C=1$ (i.e., $\alpha=f_{\rm cut}^{-2/3}$). Using equation
\ref{eq:omega}, the angle $\omega=\omega_{\rm max}$ is then a maximum given by
\begin{equation}
\omega_{\rm max} = \arctan \left\{- \frac{a_1 f_{\rm cut}^{-2/3}}{b_2+b_1 f_{\rm
cut}^{-2/3}} \right\}.
\end{equation} 
The constrained SNR is then given by
\begin{equation}\label{eq:rhoc1}
\rho_C^1 =  \frac{1}{\sqrt{2}}  \sqrt{V_0 + V_1 \cos 2\omega_{\rm
max} +V_2 \sin 2\omega_{\rm max}}\,.
\end{equation}

Finally, using the relation $V_1 \cos 2\omega+V_2\sin 2\omega =
\sqrt{V_1^2+V_2^2}\cos{\left(2\omega-\theta\right)}$, where $\tan \theta = V_2/V_1$, we
can re-write equation~\ref{eq:rho_phi0} in the general case where $0<\alpha_F<1$ by
imposing
$2\omega=\theta$, which gives
\begin{equation}\label{eq:rhoc}
\rho_C =  \frac{1}{\sqrt{2}} \sqrt{\left(V_0 + \sqrt{V_1^2+V_2^2}
\right)}\,,
\end{equation}
that is an identical expresion as in equation~\ref{eq:rhou_app} (i.e., $\rho_C= \rho_U$).
So, equation \ref{eq:wmax} is also valid, and $\alpha_F^C=\alpha_F^U$. More
details on this derivation can be found in~\cite{bcvnote}.

\section{Metric}

We can derive an expression for the match between two BCV templates
(described by equation~\ref{eq:phase}, \ref{eq:template} and \ref{eq:amplitude}).
First, we consider templates with the same amplitude function (i.e., the same
$\alpha$ and $f_{\rm cut}$ parameter). The overlap $\langle
h(\psi_0,\psi_{3}),h(\psi_0+\Delta
\psi_0,\psi_{3}+\Delta \psi_{3})\rangle$ between templates with
close values of $\psi_0$ and $\psi_{3}$ can be described (to second
order in $\Delta \psi_0$ and $\Delta \psi_{3}$) by the mismatch metric
$g_{ij}$~\cite{BCV}:
\begin{widetext}
\begin{equation}
\label{metric}
\langle
h(\psi_0,\psi_{3}),h(\psi_0+\Delta \psi_0,\psi_{3}+\Delta
\psi_{3})\rangle = 1-\sum_{i,j=0,3}g_{\rm ij}\,\Delta \psi_i \Delta \psi_j.
\end{equation}
\end{widetext}
The metric coefficients $g_{ij}$ can be evaluated analytically~\cite{BCV}, and are given
by 
 
\begin{equation}
g_{ij} = \frac{1}{2} \left[ M_{1} - M^T_2 M^{-1}_3 M_{2}\right]_{ij},\,
\end{equation}
where the ${\bf M}_{(1)\ldots(3)}$ are the matrices defined by
\bea
{\bf M}_{(1)}&=&\left[\begin{array}{cc} J(2n_0) & J(n_0+n_{3}) \\
J(n_0+n_{3}) &
J(2n_{3}) \end{array}\right]\,,\\
{\bf M}_{(2)}&=&\left[\begin{array}{cc} J(n_0) & J(n_{3}) \\ J(n_{0}-1) &
J(n_{3}-1) \end{array}\right]\,,\\
{\bf M}_{(3)}&=&\left[\begin{array}{cc} J(0) & J(-1) \\ J(-1) & J(-2)
\end{array} \right]\,,
\eea
where $n_0=5/3$ and $n_3=2/3$, and 
\begin{equation}\label{eq:moments}
J(n)\equiv\left[\int  \frac{|{\cal
A}(f)|^2}{S_h(f)}\frac{1}{f^n}df\right]\bigg/\left[\int  \frac{|{\cal
A}(f)|^2}{S_h(f)}df\right]\,.
\end{equation}

Let us emphasize the fact that the mismatch $\langle
h(\psi_0,\psi_{3}),h(\psi_0+\Delta \psi_0,\psi_{3}+\Delta
\psi_{3})\rangle$ is translationally invariant in the
$\psi_0\textrm{--}\psi_{3}$ plane, so the metric $g_{ij}$ is constant
everywhere since $J(n)$ is independent of $\psi_0, \psi_3$ parameters.

\end{appendix}

\label{theend}

\end{document}